\documentclass{iau}

\usepackage{amsmath}
\usepackage{graphicx}
\usepackage{multirow}
\usepackage{natbib}

\begin{document}

\lefttitle{Bhardwaj A. et al.}
\righttitle{Physical parameters of Cepheid and RR Lyrae stars}

\jnlPage{1}{7}
\jnlDoiYr{2021}
\doival{10.1017/xxxxx}

\aopheadtitle{Proceedings IAU Symposium} 
\editors{A. Mahabal,  C. Fluke \&  J. Mclver, eds.}

\title{Predicting Physical Parameters of Cepheid and RR Lyrae variables in an Instant with Machine Learning}

\author{A. Bhardwaj$^1$, E. P. Bellinger$^2$, S. M. Kanbur$^3$ \& M. Marconi$^1$}
\affiliation{$^1$INAF-Osservatorio Astronomico di Capodimonte, Salita Moiariello 16, 80131, Naples, Italy \\ email: {\tt anupam.bhardwaj@inaf.it} \\ 
$^2$Max Planck Institute for Astrophysics, Karl-Schwarzschild-Stra\ss e 1, 85748, Garching, Germany\\
$^3$Department of Physics, State University of New York, Oswego, NY 13126, USA\\}

\begin{abstract}
We present a machine learning method to estimate the physical parameters of classical pulsating stars such as RR Lyrae and Cepheid variables based on an automated comparison of their theoretical and observed light curve parameters at multiple wavelengths. We train artificial neural networks (ANNs) on theoretical pulsation models to predict the fundamental parameters (mass, radius, luminosity, and effective temperature) of Cepheid and RR Lyrae stars based on their period and light-curve parameters. The fundamental parameters of these stars can be estimated up to 60 percent more accurately when the light-curve parameters are taken into consideration. This method was applied to the observations of hundreds of Cepheids and thousands of RR Lyrae in the Magellanic Clouds to produce catalogs of estimated masses, radii, luminosities, and other parameters of these stars.
\end{abstract}


\maketitle

\section{Introduction}

Classical Cepheids and RR Lyrae variable stars are well known distance indicators and are also excellent tracers of young and old age stellar populations, respectively \citep{subramanian2017, beaton2018, bhardwaj2020, bhardwaj2022}. Modern pulsation codes can reproduce observed pulsation periods, light and radial velocity curves, and peak-to-peak amplitude variations for these classical pulsating stars at all wavelengths \citep{bono2000d, marconi2013, marconi2015}. In the era of large variability surveys, a quantitative comparison of the predicted and observed pulsation properties of Cepheid and RR Lyrae variables provides stringent constraints on their intrinsic evolutionary parameters, and subsequently, provides new challenges for the stellar evolution and pulsation theory \citep{bhardwaj2015, bhardwaj2017, das2018}. We aim to employ modern automated methods for comparing observed and predicted light curve structure, which is defined by its amplitude and phase parameters, of Cepheid and RR Lyrae variables and generate catalogs of physical parameters of observed variables in the Galaxy and the Magellanic Clouds.

\section{Analysis and results}

Our theoretical model grid includes 390 models with composition (Y = 0.25 and Z = 0.008) representative of Cepheids in the Large Magellanic Cloud (LMC) computed by \citet{marconi2013}, and a total of 270 models representative of RR Lyrae metallicities ranging from Z = 0.0001 to Z = 0.02 and helium abundances ranging from Y = 0.245 to Y = 0.27 \citep{marconi2015}. The pulsation models provide multiband theoretical light curves for a broad range of stellar masses, luminosity levels, and chemical composition. The observational $V-$ and $I-$band light curves were taken from the Optical Gravitational Lensing Experiment survey \citep{soszynski2018}. We use ANNs trained using {\it scikit-learn} on the grid of theoretical models to predict physical parameters of observed stars \citep[see][for more details]{bellinger2020}. The light curve structure importance was independently verified using random forests.

\begin{figure}
      \centering
        \begin{tabular}{@{}cc@{}}
    \includegraphics[scale=0.32]{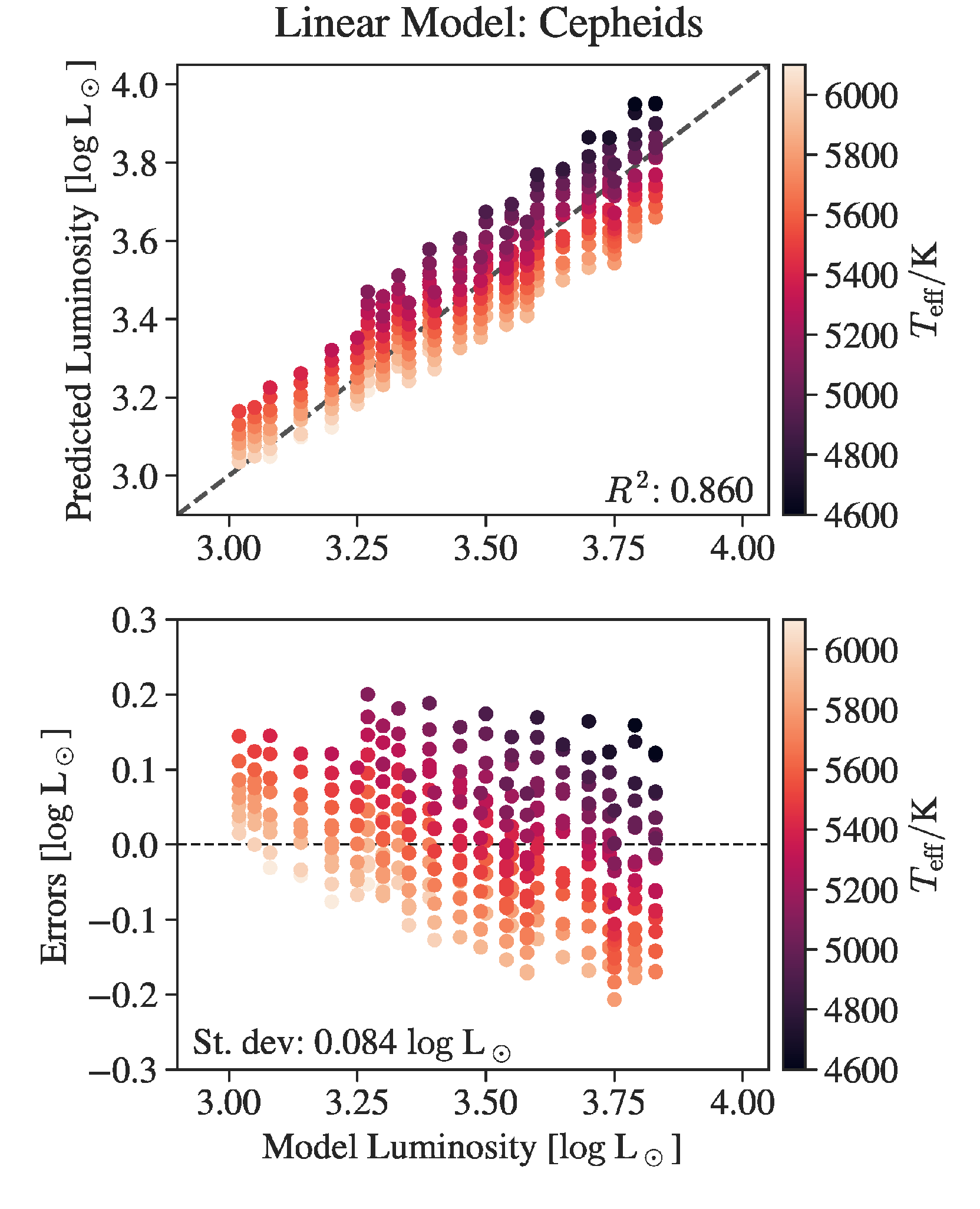} &
    \includegraphics[scale=0.32]{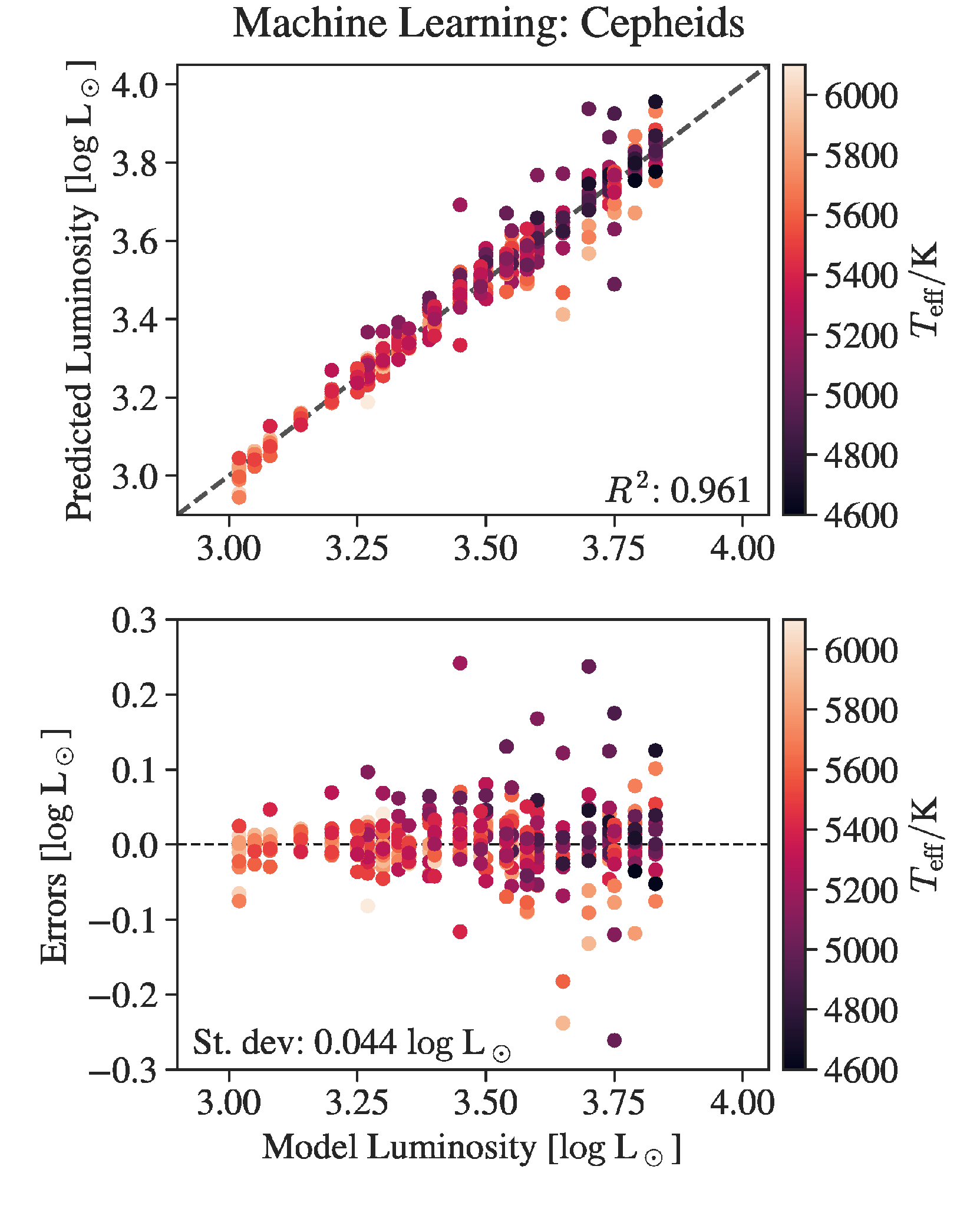}   \\
  \end{tabular}
      \caption{Predicted versus actual luminosities for estimates based on the period (left) and period and light curve parameters (right). The trends in errors (left) vanish when the light curve structure is included (right). }
\label{fig:ml_lm}
    \end{figure}

We evaluate how well the ANN can predict physical parameters based on the theoretical models and using: 1) period (linear model);  2) period and light curve structure (machine learning method). The two-fold cross validation method was used for model assessment and a quantification of improvement was provided using the standard deviation of the errors and the coefficient of variation ($R^2)$ - a higher $R^2$ is better.  In Figure~\ref{fig:ml_lm}, a significant increase in $R^2$ and a large reduction in the standard deviation of residuals is clearly seen when using the machine learning method. Similar results were also obtained for masses, radii, temperatures, absolute magnitudes, colors, and Wesenheit magnitudes \citep{bellinger2020}. 

\section{Conclusions}
While the period–mean density relation suggests that the observable period is the most important quantity in constraining the global stellar parameters of classical pulsating stars, we showed that the light curve structure plays a statistically significant part in determining these parameters. Once a smooth grid of pulsation models covering the entire parameter space is available, the catalogs of physical parameters can be instantly generated for all observed Cepheid and RR Lyrae stars with unprecedented precision.

    \bibliographystyle{mn2e}
    \bibliography{mybib_final.bib}

\begin{thebibliography}{12}
\expandafter\ifx\csname natexlab\endcsname\relax\def\natexlab#1{#1}\fi

\bibitem[{{Beaton} {et~al}\mbox{.}(2018){Beaton}, {Bono}, {Braga}, {Dall'Ora},
  {Fiorentino}, {Jang}, {Mart{\'\i}nez-V{\'a}zquez}, {Matsunaga}, {Monelli},
  {Neeley}, \& {Salaris}}]{beaton2018}
{Beaton} R.~L. {et~al.}, 2018, Space Science Reviews, 214, 113

\bibitem[{{Bellinger} {et~al}\mbox{.}(2020){Bellinger}, {Kanbur}, {Bhardwaj},
  \& {Marconi}}]{bellinger2020}
{Bellinger} E.~P. {et~al.}, 2020, MNRAS, 491, 4752

\bibitem[{{Bhardwaj}(2020)}]{bhardwaj2020}
{Bhardwaj} A., 2020, Journal of Astrophysics and Astronomy, 41, 23

\bibitem[{{Bhardwaj}(2022)}]{bhardwaj2022}
{Bhardwaj} A., 2022, Universe, 8, 122

\bibitem[{{Bhardwaj} {et~al}\mbox{.}(2017){Bhardwaj}, {Kanbur}, {Marconi},
  {Rejkuba}, {Singh}, \& {Ngeow}}]{bhardwaj2017}
{Bhardwaj} A. {et~al.}, 2017, MNRAS, 466, 2805

\bibitem[{{Bhardwaj} {et~al}\mbox{.}(2015){Bhardwaj}, {Kanbur}, {Singh},
  {Macri}, \& {Ngeow}}]{bhardwaj2015}
{Bhardwaj} A. {et~al.}, 2015, MNRAS, 447, 3342

\bibitem[{{Bono}, {Marconi} \& {Stellingwerf}(2000){Bono}, {Marconi}, \&
  {Stellingwerf}}]{bono2000d}
{Bono} G. {et~al.}, 2000, A\&A, 360, 245

\bibitem[{{Das} {et~al}\mbox{.}(2018){Das}, {Bhardwaj}, {Kanbur}, {Singh}, \&
  {Marconi}}]{das2018}
{Das} S. {et~al.}, 2018, MNRAS, 481, 2000

\bibitem[{{Marconi} {et~al}\mbox{.}(2015){Marconi}, {Coppola}, {Bono}, {Braga},
  {Pietrinferni}, {Buonanno}, {Castellani}, {Musella}, {Ripepi}, \&
  {Stellingwerf}}]{marconi2015}
{Marconi} M. {et~al.}, 2015, ApJ, 808, 50

\bibitem[{{Marconi} {et~al}\mbox{.}(2013){Marconi}, {Molinaro}, {Ripepi},
  {Musella}, \& {Brocato}}]{marconi2013}
{Marconi} M. {et~al.}, 2013, MNRAS, 428, 2185

\bibitem[{{Soszy{\'n}ski} {et~al}\mbox{.}(2018){Soszy{\'n}ski}, {Udalski},
  {Szyma{\'n}ski}, {Wyrzykowski}, {Ulaczyk}, {Poleski}, {Pietrukowicz},
  {Koz{\l}owski}, {Skowron}, {Skowron}, {Mr{\'o}z}, {Rybicki}, \&
  {Iwanek}}]{soszynski2018}
{Soszy{\'n}ski} I. {et~al.}, 2018, Acta Astron., 68, 89

\bibitem[{{Subramanian} {et~al}\mbox{.}(2017){Subramanian}, {Marengo},
  {Bhardwaj}, {Huang}, {Inno}, {Nakagawa}, \& {Storm}}]{subramanian2017}
{Subramanian} S. {et~al.}, 2017, Space Science Reviews, 212, 1817

\end{thebibliography}


\end{document}